\documentclass[8pt]{article}

\usepackage{float}
\usepackage{amsmath}
\usepackage{amssymb}
\usepackage{subfigure}
\usepackage{algorithm}
\usepackage[noend]{algorithmic}
\usepackage{graphicx}

\usepackage{hyperref} 

\usepackage{multicol}

\usepackage{cite}
\usepackage{color} 

\usepackage[labelfont=bf,labelsep=period,justification=raggedright]{caption}
\usepackage{epsfig}
\usepackage{algorithm}
\usepackage[noend]{algorithmic}
\usepackage{algorithmic}
\usepackage{graphicx}
\usepackage{subfigure}
\usepackage{amssymb}
\usepackage{amsmath}
\usepackage{stmaryrd}

\date{}

\begin{document}

\begin{flushleft}
{\Large
\textbf{Self-healing networks: redundancy and structure}
}
\\
Walter Quattrociocchi$^{\,1,2,3}$, 
Guido Caldarelli$^{\,3,4,2}$ and
Antonio Scala$^{\,4,2,3}$\\
\bf{1} Laboratory for the modeling of biological and socio-technical systems, 
Northeastern University, Boston, MA 02115 USA\\
\bf{2} LIMS the London Institute of Mathematical Sciences, 22 South Audley St 
Mayfair London W1K 2NY, UK\\
\bf{3} IMT Alti Studi Lucca, piazza S. Ponziano 6, 55100 Lucca, Italy\\
\bf{4} ISC-CNR Physics Dept., Univ. "La Sapienza" Piazzale Moro 5, 00185 Roma, Italy
\end{flushleft}

\section*{Abstract}

We introduce the concept of self-healing in the field of complex networks. Obvious
applications range from infrastructural to technological networks. By exploiting the presence of redundant links in recovering the connectivity of the system, we introduce self-healing capabilities through the application of distributed communication protocols granting  the "smartness" of the system. We analyze the interplay between redundancies and smart reconfiguration protocols in improving the resilience of networked infrastructures to multiple failures; in particular, we measure the fraction of nodes still served for increasing levels of network damages. We study the effects of different connectivity patterns (planar square-grids, small-world, scale-free networks) on the healing performances. The study of small-world topologies shows us that the introduction of some long-range connections in the planar grids greatly enhances the resilience to multiple failures giving results comparable to the most resilient (but less realistic) scale-free structures.

\begin{multicols}{2}

\section*{Introduction}

Nowadays, one of the most pressing and interesting scientific challenges deals with the analysis and the understanding of processes occurring on complex networks \cite{Arenas200893,BPPSH10,liu11,Dorogovtsev2008,Costa2007c,perra2012,barrat2008dynamical,Alvarez-HamelinFVZ12,Casteigts:2012:TVG}; one of the most important target for applying the results of such a field are real infrastructural networks.
Our society critically depends on the continuity of functioning of Physical Networked Infrastructures ($PNI$s) like power, gas or water distribution; securing such critical infrastructures against accidental or intentional malfunctioning is a key issue both in Europe and in the US \cite{CipUSgov2003,CipEUgov2006}.
While most studies have been focused on how to improve the \emph{robustness} (i.e. the capability of surviving intentional and/or random failures) of existing networks\cite{Schneider2011}, much less has been done regarding the \emph{resilience} (i.e. the capability of recovering failures). In fact, the implementation of smart (as well as economic) strategies aimed at maintaining high level of performances is a crucial issue yet to be solved. 

Most of the critical $PNI$s are very well engineered systems characterized by flow conservations (Kirchoff's laws); as a standard, these systems have been designed to be $N-1$ robust -- i.e., they should be resilient to the loss of a single component via automatic or human guided interventions. On the other hand, the constantly growing size of $PNI$s has increased the possibility of multiple failures which are not considered in such $N-1$ criteria. 
In general, implementing robustness to \textit{any sequence} of $k$ failures ($N-k$ robustness) requires an exponentially growing effort in means and investments; it is therefore viable to consider implementing systems that are \textit{on average} robust to $k$ failures.
Moreover, the continuous growth and development of $PNI$s has introduced complex connectivity (as well as dependence) patterns within their constitutive components that can often trigger systematic side effects like system-wide cascades of failures. 
Hence, our current and future $PNI$s need healing mechanisms that are able to cope with systemic effects and multiple failures in an automatic and possibly distributed way.

Healing algorithms have recently been the subject of massive investigation especially in the field of communication \cite{FlocchiniPPSW08,VelazquezS10,FlocchiniEPPS12} and wireless networks \cite{Sarma2012,Kawamura06,Murakami98,Ramchurn2012}. In particular, the strategies aimed at maintaining network connectivity assume the possibility of creating anew communications channels among the nodes of the networks, often with no constraints on the number of new connections available \cite{Pandurangan2011}. The possibility to create new links among nodes is normally not available in $PNI$s, where the links are physical (fixed in advance) and creating new links requires both time and investments.

In general, healing should be though as a constrained mechanism in which only a limited amount of resources is available; as an example, in material science new polymeric compounds are capable of self healing due to the presence of small amounts of healing agents that gets released and activated upon cracking \cite{White2001,Toohey2007}. An alternative strategy to ensure the continuity of a system is given by the redundancy of the interconnectivity of its components; for example, when a hole is punched in a leaf, the remaining vessels are capable to sustain the extra flow necessary to keep the tissues alive \cite{KatiforiPRL2010}; such redundant structures are also optimal respect to fluctuations in the flows \cite{CorsonPRL2010}. In general, technological networks can be modeled with distance-dependent constraints \cite{Barthelemy20111} that must be extended to more sophisticated constraints like the shape of the spectrum in the case of power grids \cite{Wang2010}.

In this paper we introduce a novel strategy to enhance the response and resilience of networked infrastructures by studying the effect of {\em randomness} and {\em redundancy} in their connectivity patterns.

As mentioned above, such an approach aims at studying the impact and the feasibility of cost-effective strategies to improve the resilience of physical infrastructure. 
In fact, we introduce a healing strategy based on the activation of fixed redundant resources (backup links) and study the resilience of the networks to multiple failures. Since the presence of backup links is customary in technological networks and our strategy can be implemented via routing protocols, our self-healing procedure is within the reach of current technology. 

\section*{Results}

\subsection*{Model}

In our scenario the system is assumed to be a network that distributes some utility (like water, power, gas, oil) following Kirchoff (continuity) laws. For sake of simplicity, we will consider a single node to be the source of the quantity to be distributed on the network. Moreover, at each instant of time, the topology of the network distributing the utility (the \emph{active tree}) is assumed to be a tree. 
In fact, such a structure meets the infrastructures' managers needs -- i.e., to measure (for billing purposes) in an easy and precise way how much of a given quantity is served to any single node of the network.
Finally, as a further simplification we will not take into account the magnitudes of flows -- i.e., all links and sources are assumed to have infinite capacity -- but we will focus on maximizing the connectedness of the system in order to serve as many nodes as possible.

In order to implement our strategy and its self-healing capabilities, we consider the presence of {\em dormant} backup links -- i.e., a set of links that can be switched on. 
Nodes are assumed to be able to communicate with their neighbours by means of a suitable distributed interaction protocol with a limited amount of knowledge: the set of neighbouring nodes connected either via active or via dormant links. 
Then, when either a node or a link failure occurs, all the nodes below the failure will disconnect from the active tree and become unserved. Such unserved nodes can now try to reconnect the active tree by waking up through the protocol some dormant backup links. Such a process will reconstruct a new active-tree that can restore totally or partially the flow, i.e. heal the system. 

In order to identify the system's properties that are able to maximize the fraction of service ($FoS$)  -- i.e, the fraction of served nodes -- we study the effects of the backup links (redundancy) disposed according to different connectivity patterns -- i.e, different underlying networks on top of which the active tree is build -- with respect to multiple random failures.
We start our investigation by focusing on the case which best resembles the actual situation of $PNI$s -- i.e. nodes disposed over a grid. Then, we stress the role of the networks' connectivity patterns by using small-world and scale free networks as underlying topology. A more formal description of the model is provided in the section~\nameref{sec:methods}.

\subsection*{Materials}

In order to stress the peculiarities of different network structures, we generate class of graphs with different connectivity patterns (see \nameref{sec:methods} ): the planar square grid ($SQ$), the scale-free ($SF$) topology of Barabasi-Albert \cite{BarabasiAlbert} and the small world ($SW$) of Watts and Strogatz\cite{WS98}.
For each of these networks, the source node -- i.e, the root of the oriented active tree - is chosen at random within all the nodes of the underlying network.  The only exception is the case of the $SF$ networks where we use, according to the preferential attachment principle, the natural choice of having  the node with the highest number of neighbours (the central hub) as the source. 

For each kind of network structures, we generate the associated random spanning trees \cite{Wilson96} (section~\nameref{sec:methods}). We take such spanning trees as the initial configuration of our model distribution networks. The links not belonging to the spanning trees form the set of the possible backup links of our system; among such links, we choose a random fraction $r$ of \emph{dormant} links that can be used to heal the system. We then simulate the occurrence of uncorrelated multiple failures by deleting at random $k$ links of the initial active tree. Notice that link failures are the most general ones, as a node failure is equivalent to the simultaneous failure of all its links. 

\begin{figure}[H]
 \centering
	\subfigure[\textbf{Initial configuration}]
     {\includegraphics[width=.25\textwidth]{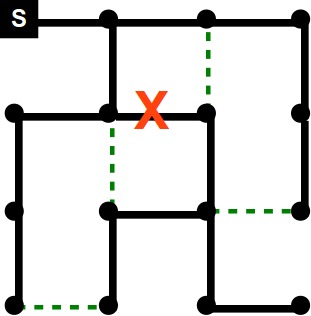} \label{subfig:Tree}} 
   \hspace{.05\textwidth}
   \subfigure[\textbf{Link failure}]
    {\includegraphics[width= .25\textwidth]{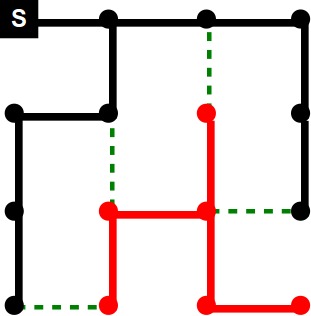} \label{subfig:LinkFail}} 
   \hspace{.05\textwidth}
     \subfigure[\textbf{Recovered network}]
       {\includegraphics[width=.25\textwidth]{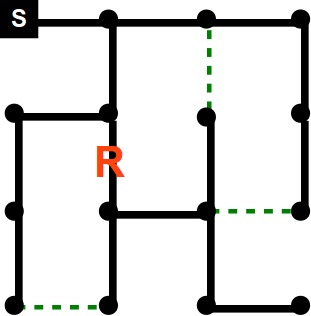} \label{subfig:LinkRecover}} 
   \caption{
(Color online) Example of healing after single link failure. Notice that failure of a single node can be modelled as the failure of all its links; hence, multiple links failure are the more general event to be considered. 
(Left Panel) In the initial state, the source node (filled square, upper left corner) is able to serve all $16$ nodes through the links of the active tree. The $4$ dashed lines (green online) represent dormant backup links that can be activated upon failure. The redundancy of the system is $p=4/9$ as at most $9$ backup links can be present. The link marked with an \textcolor{red}{\textbf{X}} is the one that is going to fail.
(Central panel) A single link failure disconnects all the nodes of a sub-tree; in the example, a sub-tree of $6$ nodes (red online) is left isolated from the source -- i.e., the system has a damage $\Delta=6$).
(Right Panel) By activating a single dormant backup link, the self-healing protocol has been able to recover connectivity for the whole system, in this case bringing back the number of served nodes at its maximum value $16$. The link that has recovered the connectivity is marked with an \textcolor{red}{\textbf{R}}.
   }
\label{fig:HealLinkFailure}
\end{figure}

Our self-healing algorithm is a routing protocol (see section~\nameref{sec:methods}) whose goal is to reconstruct the maximum spanning tree connected to the source after that a failure has occurred; in doing so, we use both the survived links of $T$ and the dormant links $D$; fig.\ref{fig:HealLinkFailure} illustrates such procedure. We indicate with $FoS$ the fraction of nodes connected to the source after the recovery.

\end{multicols}

\begin{multicols}{2}

\subsection*{Results on different networks}

In order to test the performances of our healing algorithm to failures in terms of fraction of service provided after the active tree restoration, we simulate the model for increasing number of failures. Recalling that each failure causes a cascade -- i.e, each node of sub-tree served by the broken link is unserved -- we investigate the role of redundancy $r$ on different topologies. 

\begin{figure}[H]
\centering
\subfigure[square grid]
     {\includegraphics[width=.45\textwidth]{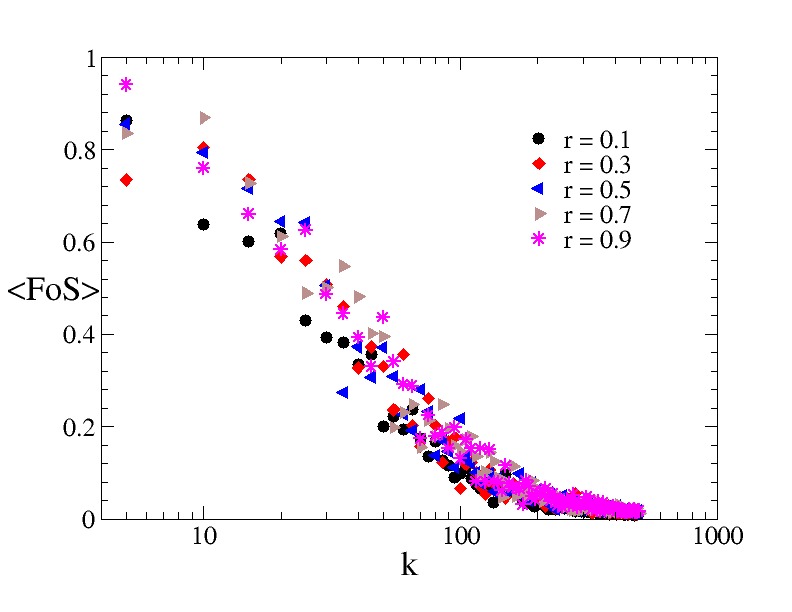} \label{subfig:SQ10k}} 
   \hspace{1mm}
\subfigure[scale free]
     {\includegraphics[width=.45\textwidth]{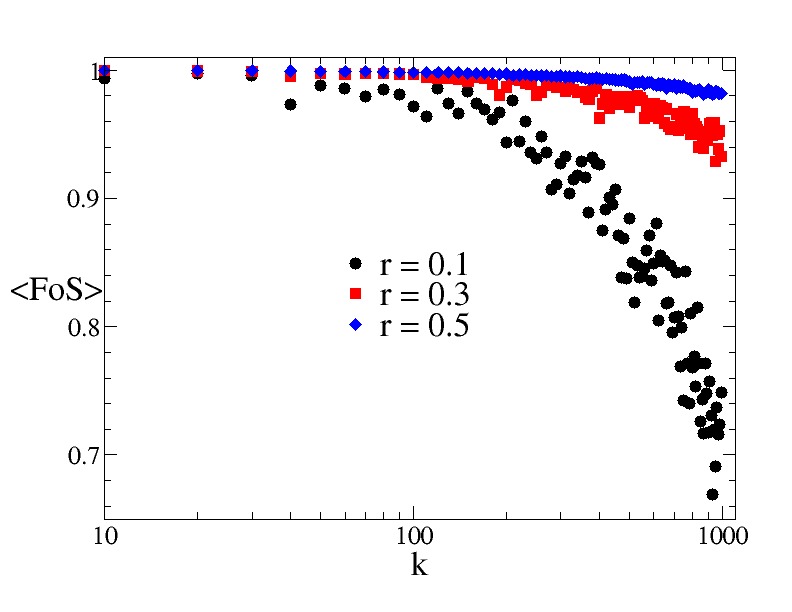} \label{subfig:SF10k}} 
   \hspace{1mm}
\subfigure[small world]
     {\includegraphics[width=.45\textwidth]{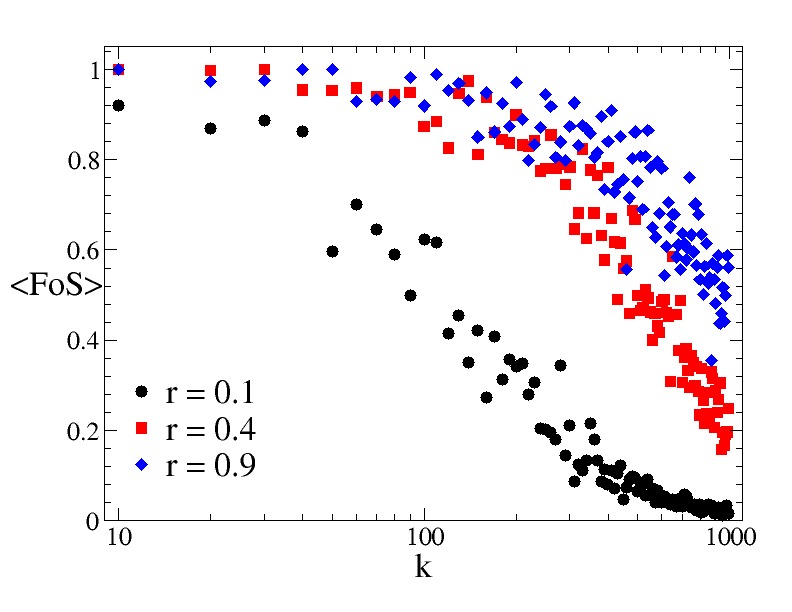} \label{subfig:SW10k}} 
\caption{ 
(Color online) Self-healing results for networks of size $10^4$. 
Panel (a): distribution networks based on square grids . The average fraction $\langle FoS\rangle$ of nodes that the self-healing protocol is able to restore decreases with the number of faults $k$ with no relevant dependency on the redundancy; results are shown for a $10^4$ nodes network.
Panel (b): distribution networks based on Barabasi-Albert networks. The average fraction of nodes$\langle FoS\rangle$ of served nodes is plotted against the number of failures $k$. Even for a low $10\%$ redundancy ($r=0.1$), the system can almost totally heal after sustaining $k\sim 4\times 10^2$ failures; as a comparison, for the same number of failures square grids loose $\sim 90\%$ of the nodes.
Panel (c): distribution grids based on small-sworld networks obtained by rewiring a fraction $p=0.2$ of links. The average fraction of nodes$\langle FoS\rangle$ of served nodes is plotted against the number of failures $k$. At difference with square grids and scale free networks, the restored fraction of service $FoS$ shows a marked dependency upon the redundancy parameter $r$. Similar results are obtained for $p=0.1$ and $p=0.3$.
}
   \label{fig:Grids10k}
\end{figure}

We start our study by addressing planar square grid ($SQ$) networks that are the most similar to the real physical networked infrastructures. In the first scenario, we generate spanning trees on a square grids; figure \ref{subfig:SQ10k} shows the variation of the restored $FoS$ respect to the number of failures $k$ for different redundancies $r$s. For square grids, we do not observe any relevance of the redundancy on the $FoS$; this means that a very small fraction backup links ($r=0.1$, i.e. $10\%$) already suffice to attain the maximum resilience. 

The situation is completely different when the underlying topology is a scale-free network generated through the Barabasi-Albert model \cite{BarabasiAlbert}. A widely diffused property of real networks is that the connectivity pattern follows a scale-free power-law distribution \cite{Vesps2001,Caldarelli2007,Takaaki2012}. This feature has been found to be a consequence of the so called {\em preferential attachment} -- i.e networks expand continuously through the addition of new vertices which attach preferentially to already well connected nodes. Although technological networks do not show power law degree distributions due to economic and spatial constraints\cite{Amaral00}, we choose to investigate $SF$ networks for their marked robustness upon random failures \cite{Bollobas2004}. 
For $SF$ networks, it is natural to choose the node with the highest degree (the hub) as the source. The quality of service restored by our self-healing algorithm on $SF$ networks is shown in Figure \ref{subfig:SF10k}. As expected, we find that $SF$ networks can easily recover all the nodes even for low redundancies. Such error tolerance comes at a high price of being extremely vulnerable to \emph{node} targeted attacks: isolating the hub disconnects the whole system. High error tolerance and targeted attack vulnerability are indeed generic properties of $SF$ networks \cite{albert2000}.

\end{multicols}
\begin{multicols}{2}

We then consider the case of small-world ($SW$) networks \cite{WS98}. In the case of technological networks, small world networks are important as they can show the effects of introducing long-range links in a planar topology. Starting from an initial graph (planar square grids in our case), we rewire with a probability $p$ a link with a randomly selected node; in this way we can interpolate from the case of $SQ$ networks ($p=0$) to the case of a random graph ($p=1$). As in the case of simple percolation \cite{NewmanPRE1999}, the rewiring procedure introduces some \emph{long range} links -- i.e., between distant nodes on the square grid) that improve the robustness to random failures.

In order to understand the role of the connectivity pattern we study our model on different $SW$ networks with different rewiring probabilities.
In Figure \ref{subfig:SW10k} we show the performances of our self-healing strategy with respect to an increasing number of failures. We see that a higher rewiring probability increases the number of served nodes after the restoration through the backup network; such a peculiarity shows up even if the clustering within neighbouring nodes (normally associated to a local robustness against failures) decreases; therefore long range links increase the possibility of the network staying connected even after multiple failures.

\begin{figure}[H]
 \centering
 \includegraphics[width=.75\columnwidth]{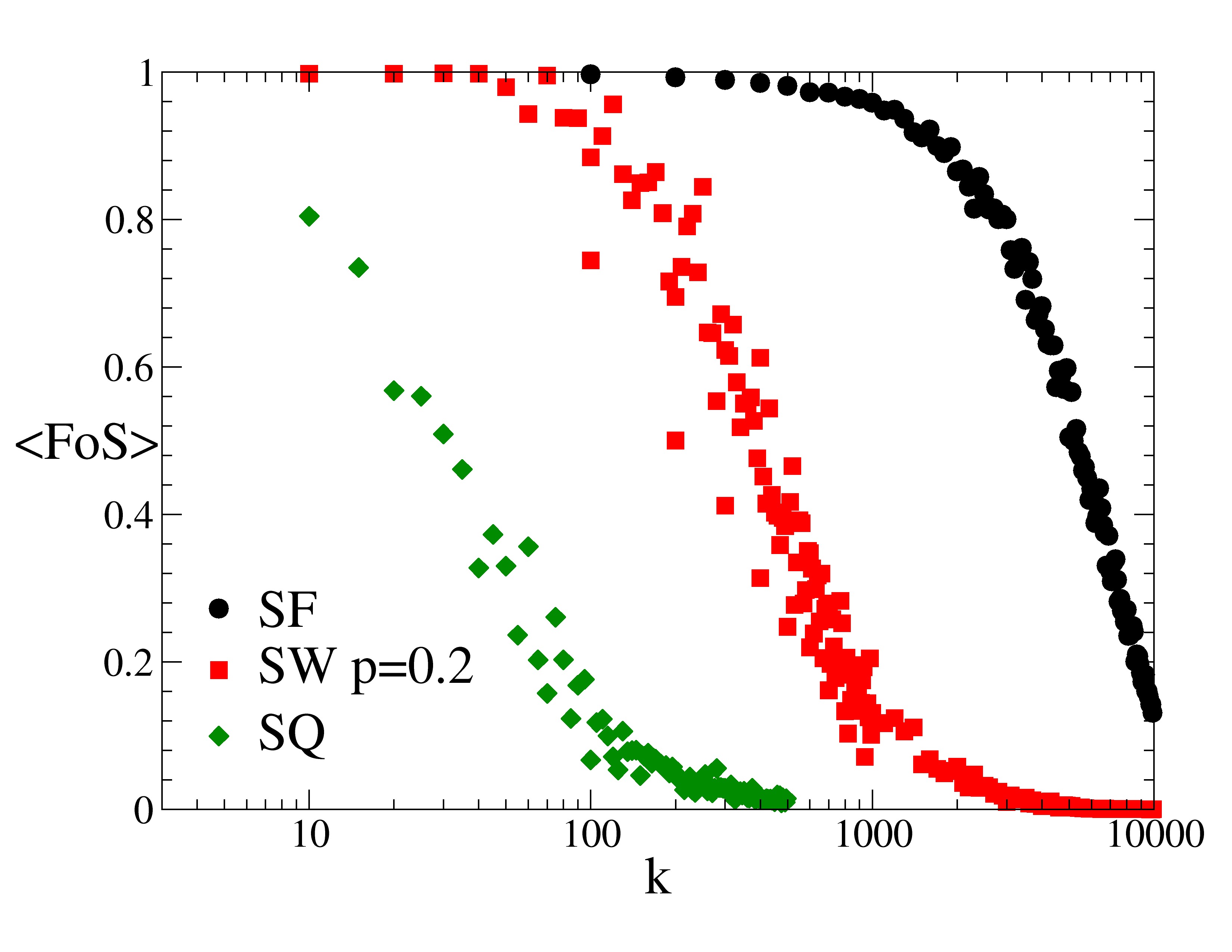}
   \caption{(Color online) Comparison among different network structures. Here we show the performances of our self-healing algorithm with respect to the quality of service for increasing number of removed links  with the redundancy $r$ fixed to $0.3$; for $SW$ networks, the rewiring probability is $p=0.2$. The average fraction of nodes$\langle FoS\rangle$ of served nodes is plotted against the number of failures $k$.}
\label{fig:Healing43topologies}
\end{figure}

Finally, we compare in fig.\ref{fig:Healing43topologies} the effectiveness of the self-healing protocol across different strategies. While distribution grids based on the $SF$ topology are the more robust, they should be disregarded when considering the case of technological networks since economic and geometric constraints make $SF$ networks unfeasible on planar topologies. 

\end{multicols}
\begin{multicols}{2}

\section*{Discussion}

In this paper we have introduced a minimal procedure of self-healing in networks. Such procedure exploits the presence of redundant edges to recover the connectivity of the system. 
Our scenario is inspired by real-world distribution networks that are, often for economic reasons, tree-like and in the meantime are also often provided with alternative backup links that can be activated in case of malfunctioning; as an example, this is the case for low-voltage distribution networks \cite{ENELprivcom}. 

Our model, albeit schematic, is realistic in the sense that it could be readily and easily implemented with the current technologies. In fact, routing protocols represent a vast available source of distributed algorithms able to maintain the connectivity of a system. Therefore, our scheme could be implemented by coupling an ICT network to current infrastructures. Our case is an example in which interdependencies enhance the resilience instead of introducing catastrophic breakdowns \cite{BPPSH10}.

We have studied the performances of our procedure varying the redundancy on different network substrates -- i.e, planar square grid, scale-free and small-world networks. Within our model, we have shown that distribution networks akin to real world ones - i.e, based on $SQ$ topologies - to be the less resilient to failures. In facts, as expected the most robust networks are based on the $SF$ topology that is unrealistic for technological networks. A further direction of study would be to consider the effects of more detailed structural characteristics on the dynamics of the system \cite{DAgostinoEPL2012}.

Our results on $SW$ topologies hint that a very effective strategy to strengthen planar networks is to add long range links. The feasibility of such a strategy depends on cost-benefit analysis about the implementation of physical long-range links in $PNI$s. 

While our minimal model considers only the connectivity of the system, it can be easily expanded to take account of the magnitude of the flows: in fact, routing algorithms can account for both the capacity of the links and dynamically swap re-routing of flows. 
Our model easily allows also for {\em cold starts} -- i.e., for situations in which the network has shut down due to some major events (like a black-out) \cite{Sudhakar2011}. This is an important issue as one of the most time (and money) consuming activity after a major event is the restoring of the functionality of the network.

In this paper, we have considered only the single source case. Next step is to consider a network served by multiple sources. The possibility of separating a networks in separate islands is a well know strategy to mitigate cascade failures in power grids; for electric distribution separating the system in trees would solve the problem of {\em who is serving who} that appears as soon as more competitors share the same physical line in bringing power to their customers \cite{LiGLOBECOM2010}. Moreover, the possibility for the system of dynamically separating in time-varying trees would allow for introducing a commodity market based on real-time economic competition among the owners of the sources. This further goal is not yet within the reach of current routing protocols and should be further investigated if we want to have grids that are smart not only for their ability to self-repair but also in optimizing consumptions and prices.

\section*{Methods}\label{sec:methods}

\subsection*{The Self-Healing procedure}

We consider an abstract model of a physical networked infrastructure ($PNI$) described by the quadruple $\left(V,v_S,E_A,E_D\right)$. Here $V$ are nodes of the network, $v_S\in V$ is the source node, $E_A$ is the set of active links among the nodes and $E_D$ denotes the set of dormant links that can be activated in order to heal system failures by re-connecting nodes. 
A node is considered to be served if it is connected to a source through a path of active links; all the nodes in $V$ are initially connected to the source. As the basic metric for any quality of service ($QoS$) assessment, we consider the fraction of served nodes $FoS$ counting the number of nodes in the active graph -- i.e connected to $v_S$.

More formally, in the initial configuration, the graph $T=\left(V,E_A\right)$ is an instance of $R_{T}(G)$ -- i.e, the set of all the random spanning tree of the graph $G=\left(V,E_A \cup E_D\right)$ with $v_S$ $\in G$. Thus, $T$ before the failures has $|V|$ (active) nodes and $\vert E_A\vert = V-1$ links among them. 

We then consider the occurrence of multiple link failures. A $k$-failure is a subset $E_F \subset E_A$ of $k$ links chosen at random. The system right after a failure is described by the graph $G'=\left(V,E_A-E_F\right)$ and by the set $E_D$ of dormant links available for the healing. A healing protocol is any algorithm that finds the maximal tree $T'$ of $G'$ containing the source $v_S$ by activating (waking up) a subset $E_W \subset E_D$ of dormant edges. 
For robustness, we will assume that nodes have only a \emph{local} knowledge of the networks -- i.e., only about the state (active, dormant or failed) of their incoming links. To build the maximal connected tree, nodes communicate with their neighbors via a suitable distributed protocol allowing fault nodes to join the active network by activating {\em dormant} edges. 
In other words, nodes are endowed only with the minimal requirements of routing needed to reconstruct a spanning tree \cite{Santoro2006}. In this paper, we have applied the following simple  distributed algorithm to implement self-healing:

\begin{algorithmic}
\STATE $U \leftarrow $ unserved nodes
\STATE $V' \leftarrow V-U $
\STATE $E'_A \leftarrow (E_A-E_F) \cap (V \times V)$
\STATE $E'_D \leftarrow (E_D-E_F) \cup (E_A - V\times V) $
\REPEAT
\FORALL{$v \in U$}
\STATE choose a random neighbor $a(v)$ connected to $V$ through any edge of $E'_D$
\ENDFOR
\FORALL{$v \in U$}
\IF{$a(v)\neq\emptyset$}
\STATE $V' = V' + \lbrace v \rbrace$
\STATE $U = U - \lbrace v \rbrace$
\STATE $E'_A = E'_A + \lbrace \left( u(v), v \right) \rbrace$
\STATE $E'_D = E'_D - \lbrace \left( u(v), v \right) \rbrace$
\ENDIF
\ENDFOR
\UNTIL{$V'$ and $U$ are disconnected or $U=\emptyset$}
\STATE $E'_D \leftarrow E_D' \cap V'\times V' $
\RETURN $T'=\left(V',v_S,E'_A,E'_D\right)$
\end{algorithmic}

By definition, the nodes in $T'$ are the set of served nodes $V'$. 
Notice that the state $\left(V',v_S,E_A+E_W-E_F,E_D-E_W\right)$ still describes a $PNI$; therefore, we can in general describe the state of the system at time $t$ by the quadruple $\left(V(t),v_S,E_A(t),E_D(t)\right)$ and the sequence of time failures between time $t$ and $t+1$ by $E_F(t)$. A general representation of such a process can be given in terms of time varying graph \cite{Casteigts:2012:TVG}.

\subsection*{Simulations}

\begin{figure}[H]
 \centering
\subfigure[Grid]
     {\includegraphics[width=.26\textwidth]{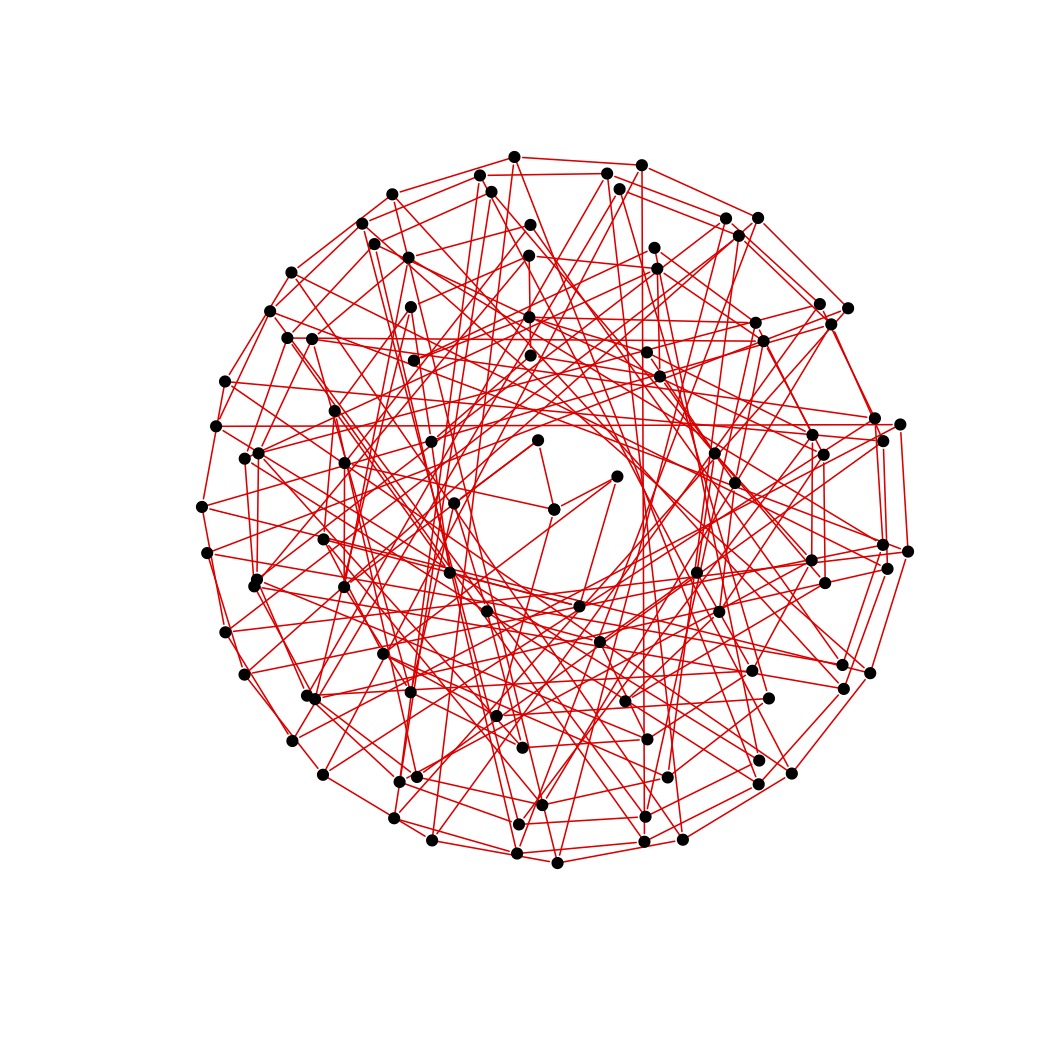} \label{subfig:SQ}} 
   \hspace{1mm}
   \subfigure[Watts-Strogatz]
    {\includegraphics[width= .26\textwidth]{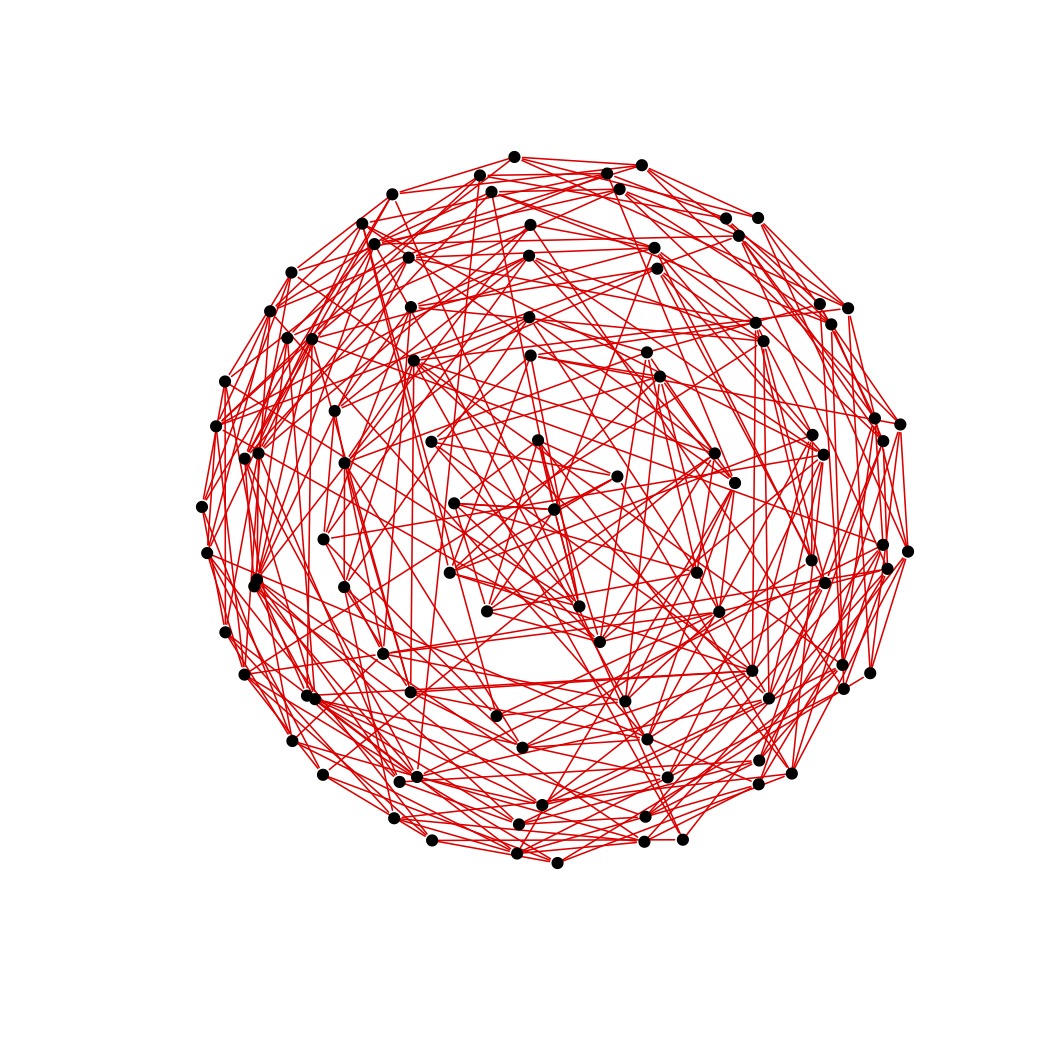} \label{subfig:WS}} 
      \hspace{1mm}
     \subfigure[Barabasi-Albert]
       {\includegraphics[width=.26\textwidth]{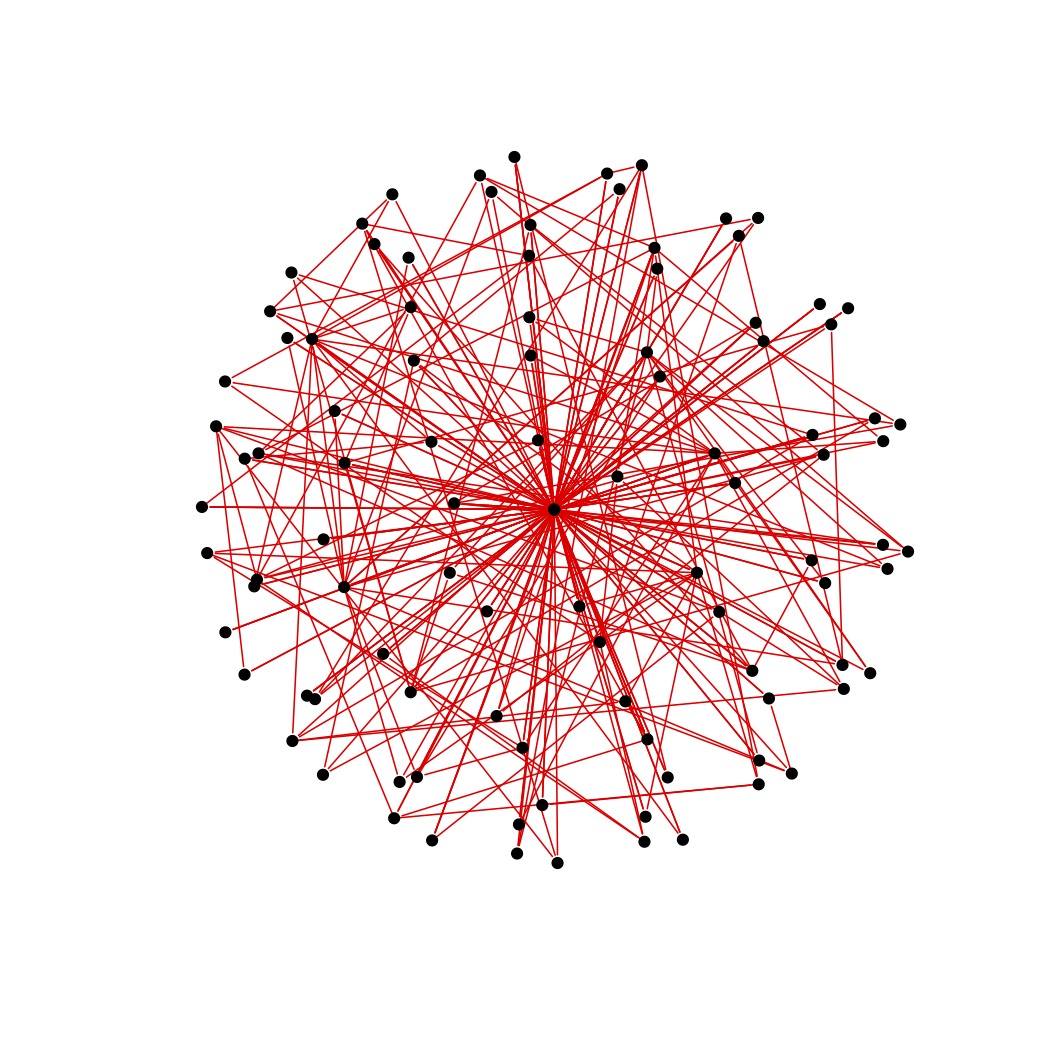} \label{subfig:BA}} 
      \subfigure[Grid]
     {\includegraphics[width=.26\textwidth]{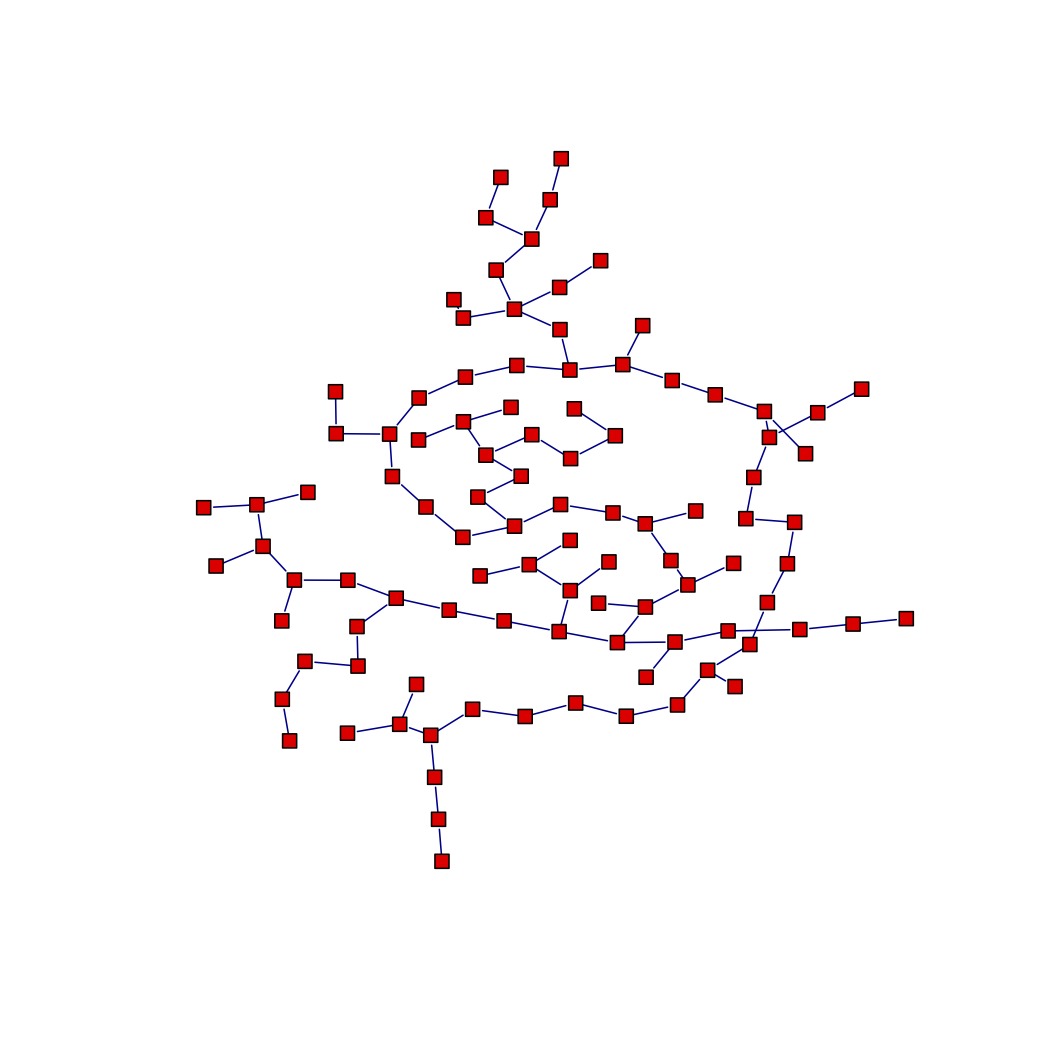} \label{subfig:SQtree}} 
   \hspace{1mm}
   \subfigure[Watts-Strogatz]
    {\includegraphics[width= .26\textwidth]{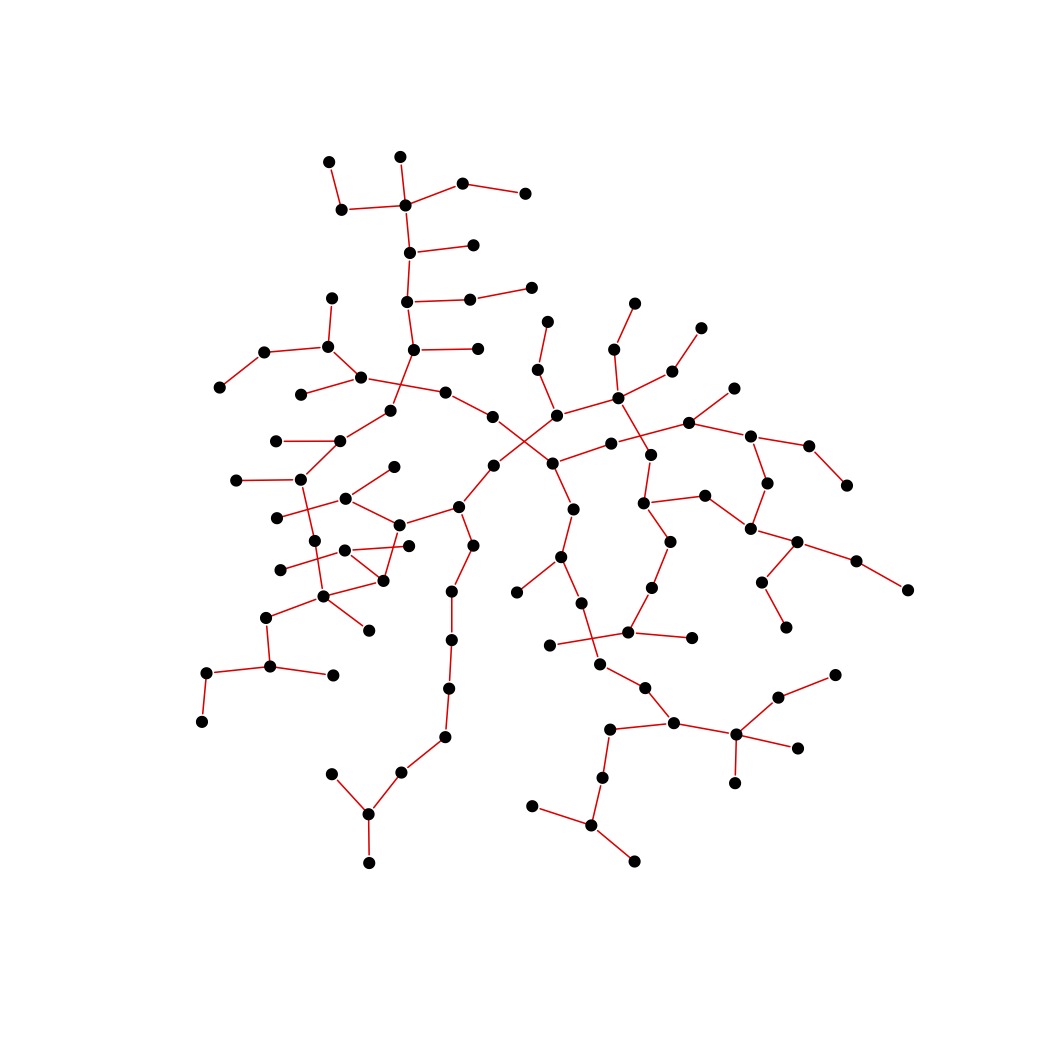} \label{subfig:WStree}} 
      \hspace{1mm}
     \subfigure[Barabasi-Albert]
       {\includegraphics[width=.26\textwidth]{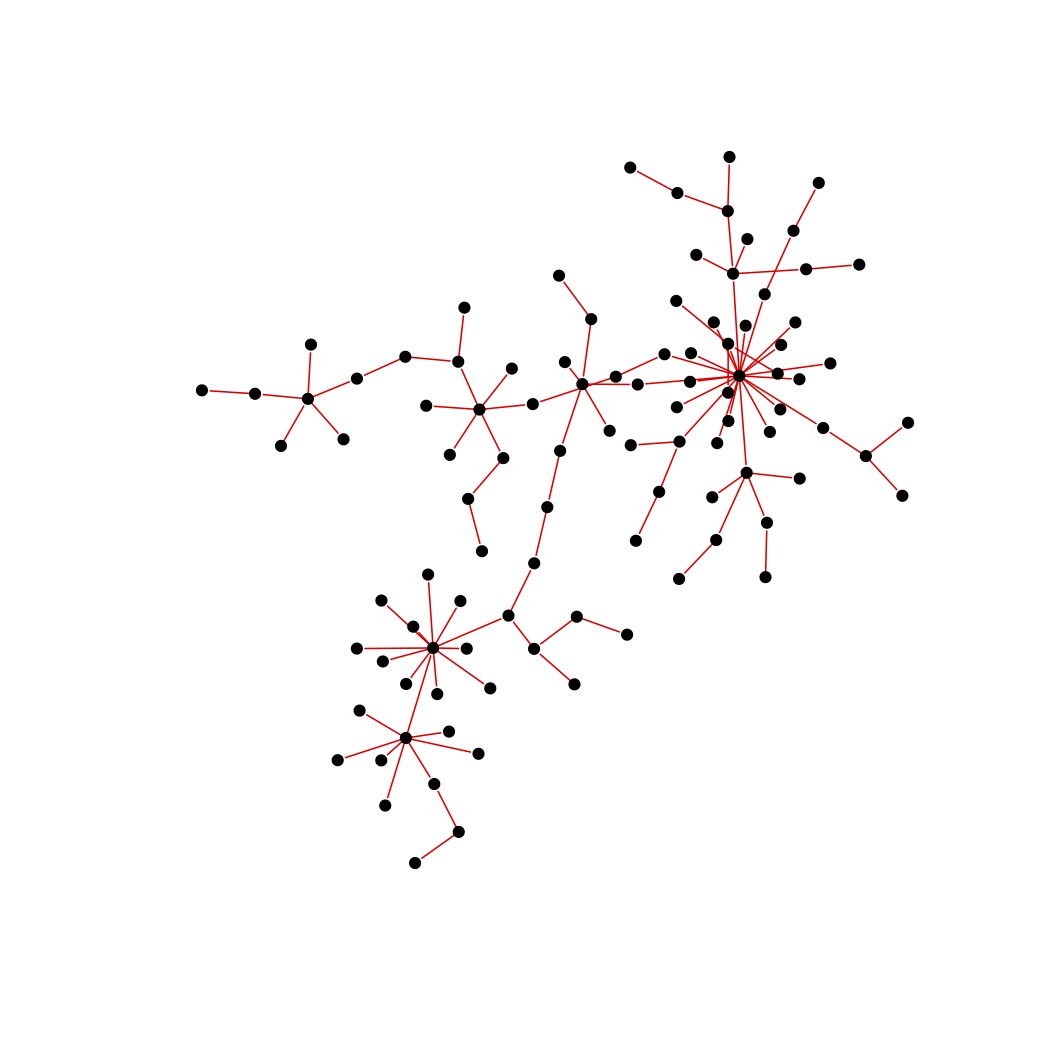} \label{subfig:BAtree}} 
      \hspace{1mm}
   \caption{(Color online) Different network topologies. Upper panels, from left to right: planar square grid (SQ), Watts-Strogatz small world network (SW) and Barabasi-Albert scale-free network (BA). Lower panels: random spanning trees associated with the related underlying topologies in the upper panels.}
   \label{fig:GraphsAndTrees}
\end{figure}

We perform simulations on different grids (fig.\ref{fig:GraphsAndTrees} -- upper panel) averaging over different possible initial configuarions.
To generate a random spanning tree $T$ associated to a graph $G$ (fig.\ref{fig:GraphsAndTrees} -- lower panel), we apply the exact algorithm of Wilson \cite{Wilson96} that samples uniformly the elements of $R_{T}(G)$. Such spanning trees are taken as the initial configurations for our model distribution networks. 
The links of the graph $G$ that do not belong to the initial configuration $T$ form the set $E_B$ of the possible backup links of our system; of such links, only a subset $E_D$ (the \emph{dormant} links) can be used to heal the system. The fraction $r=\vert E_D \vert / \vert E_B \vert$ of such dormant links characterizes the redundancy of the system: for $r=0$ there are no links in $E_D$ and any failure splits the tree, while for $r=1 $ any of the links of $G$ can be used to recompose the system.

\end{multicols}
\begin{multicols}{2}

We analyze the response of the system to $k$-failures. Classically, engineered systems are built $N-1$ robust, i.e. they can survive to the single failure of any of their components. While checking $N-1$ robustness corresponds to checking all the $N$ possible single failures, checking the $N-k$ robustness requires to consider $N!/(N-k)!\sim N^k$ cases. Therefore, checking the $N-k$ robustness is infeasible even for modest values of $k$ due to the combinatorial explosion of the number of possible cases. Thus, we choose to assess on probabilistic ground whether a system would be able to sustain $k$ failures by a Monte Carlo investigation of the space of possible failures.
 
Service operators are interested in maintaining their service level agreements (contracts) with their customers; to such an aim, customers must in first pace remain connected to the services. 
Therefore, we calculate the average fraction of served customers $FoS$ after the occurrence and the healing of $k$ random failures. To do so, we choose at random $k$ different links on the service tree and delete them; after that, we apply the self healing procedure; finally, we calculate the $FoS$. Such procedure is repeated until the relative error of the average $FoS$ is small enough (less than $5\%$). 
As an example, for a grid $100 \times 100$ of $10^4$ nodes, we must average over $100$ sets of random failures to attain the desired accuracy. Moreover, to average out the different characteristics of the initial configurations, we repeat the procedure over $100$ different independently generated initial configurations.

Notice that in our case it is more correct to speak about $N-k$ \emph{resilience}, since we don't consider whether the system is \emph{robust} to $k$ failures (i.e. whether it still functioning after $k$ failures), but if it can \emph{recover} from $k$ failures.

\section*{Acknowledgements}

We thank US grant HDTRA1-11-1-0048, EU FET Open project PLEASED nr.296582, CNR-PNR National Project \textquotedblright{}Crisis-Lab\textquotedblright{} and EU FET project MULTIPLEX nr.317532. 
The contents of the paper do not necessarily reflect the position or the policy of funding parties. 
AS thanks Claudio Mazzariello 
for very useful discussions on routing algorithms, Thomas Fink for suggesting the focus on Kirchoff networks, Stefano Sello and Matteo Cant\'{u} 
for discussions on the topology of Italian distribution networks and Michele Festuccia 
for pointing out the technological feasibility of our approach. 
WQ thanks Alan Advantage Italy and Alan Advantage U.S. for insightful technical discussions.

\section*{Additional Information}
\textbf{Competing financial interests} The authors declare no competing financial interests.

\bibliographystyle{unsrt}



\section*{Contributions}
AS and WQ equally contributed to the simulations and the data analysis. 
AS, WQ and GC equally contributed to the writing of the paper.

\end{multicols}

\end{document}